\documentclass{aa}  
\usepackage{graphicx}
\usepackage{amssymb}

\begin{document}
\title{The performance of the blue prime focus Large Binocular Camera
at the Large Binocular Telescope}

\author{
          E. Giallongo \inst{1}
          \and
          R. Ragazzoni \inst{2}
          \and
          A. Grazian \inst{1}
          \and
          A. Baruffolo \inst{2}
          \and
          G. Beccari \inst{3}
          \and
          C. De Santis \inst{1}
          \and
          E. Diolaiti \inst{3}
          \and
          A. Di Paola \inst{1}
          \and
          J. Farinato \inst{2}
          \and
          A. Fontana \inst{1} 
          \and
          S. Gallozzi \inst{1}
          \and
          F. Gasparo \inst{4}
          \and
          G. Gentile \inst{2}
          \and
          R. Green \inst{5}
          \and
          J. Hill \inst{5}
          \and
          O. Kuhn \inst{5}
          \and
          F. Pasian \inst{4}
          \and
          F. Pedichini \inst{1}
          \and
          M. Radovich \inst{6}
          \and
          P. Salinari \inst{7}
          \and
          R. Smareglia \inst{4}
          \and
          R. Speziali \inst{1}
          \and
          V. Testa \inst{1}
          \and
          D. Thompson \inst{5}
          \and
          E. Vernet \inst{7,8}
          \and
          R. M. Wagner \inst{5}
}

\offprints{E. Giallongo, \email{giallongo@mporzio.astro.it}}

\institute{INAF - Osservatorio Astronomico di Roma, Via Frascati 33,
I--00040, Monteporzio, Italy
\and INAF - Osservatorio Astronomico di Padova, vicolo dell'Osservatorio 5,
I--35122 Padova, Italy
\and INAF - Osservatorio Astronomico di Bologna, Via Ranzani 1,
I--40127 Bologna, Italy
\and INAF - Osservatorio Astronomico di Trieste, Via G. B. Tiepolo 11,
I--34131 Trieste, Italy
\and Large Binocular Telescope Observatory, University of Arizona, 933 N.
Cherry Ave., Tucson, AZ 85721-0065
\and INAF - Osservatorio Astronomico di Capodimonte, via Moiariello 16,
I--80131, Napoli, Italy
\and INAF - Osservatorio Astronomico di Arcetri, Largo E. Fermi 5,
I--50125, Firenze, Italy
\and European Southern Observatory, Karl-Schwarzschild-Str. 2,
D--85748, Garching, Germany
}

   \date{Received August 2, 2007; accepted January 8, 2008}

   \authorrunning{E. Giallongo et al.}
   \titlerunning{The performance of the blue prime focus LBC}

 
\abstract
{}
{
We present the characteristics and some early scientific results
of the first instrument at the Large Binocular
Telescope (LBT), the Large Binocular Camera (LBC). Each LBT telescope
unit will be equipped with similar prime focus cameras.  The blue
channel is optimized for imaging in the UV-B bands and the red channel
for imaging in the VRIz bands.  The corrected field-of-view
of each camera is approximately 30 arcminutes in diameter, and the chip area is
equivalent to a $23\times 23$ arcmin$^2$ field. In this paper
we also present the commissioning results of the blue channel.
}
{
The scientific and technical performance of the blue channel was assessed
by measurement of the astrometric distortion, flat fielding, ghosts, and
photometric calibrations. These
measurements were then used as input to a data reduction pipeline
applied to science commissioning data.
}
{
The measurements completed during commissioning show that the
technical performance of the blue channel is in agreement with original
expectations. Since the red camera is very similar to the blue
one we expect similar performance from the commissioning that will be
performed in the following  months in binocular configuration. Using
deep UV image, acquired during the commissioning of the blue camera, we
derived faint UV galaxy-counts in a $\sim 500$ sq. arcmin. sky area to
U(Vega)$=26.5$. These galaxy counts imply that the blue
camera is the most powerful UV imager presently available and in the
near future in terms of depth and extent of the field-of-view. We
emphasize the potential of the blue camera to increase the robustness of
the UGR multicolour selection of Lyman break galaxies at redshift
$z\sim 3$.
}
{}


\maketitle
%

\section{Introduction}

The sensitivity of an optical system depends on a combination of the
aperture and field-of-view (FoV).

The imaging capabilities of existing or planned facilities are often
limited by practical constraints. When the large collecting area of a
telescope allows detection of faint sources, the field-of-view is
typically less than $7\times 7$ square arcminutes, and the UV sensitivity is
low. Alternatively, wide-field imaging cameras onboard smaller
telescopes are optimized to target brighter sources over a larger
field-of-view (i.e. MegaCam at CFHT, \cite{boulade}), and are unable
to detect sources of faint magnitudes ($\sim 28$) in particular in the UV.

For these reasons an imager with a large FoV at an 8m class telescope
is of fundamental importance to address the presently still open
problems in stellar and extragalactic astronomy.
The best example is the prime focus camera at the 8m Subaru telescope,
Suprimecam (\cite{subaru}). This imager is fast and has a FoV of
$34\times 27$ arcmin$^2$. Common science projects that have utilized
this imager to date are the search of very high redshift
galaxies, the study of the formation and evolution of galaxies, the
investigation of the structure of the Universe, and the search for
Kuiper Belt objects in the Solar system.
The optical corrector cannot, however, simultaneously correct
radiation of all wavelengths from UV to I-band. Due to this practical
limitation, and to its low sensitivity in the blue band, Suprimecam
does not provide imaging in the UV.

At the end of the 1990s, it became clear that the binocular
configuration of the Large Binocular Telescope (LBT) (\cite{hill}),
coupled with its mechanical design, provided a unique opportunity to
justify a double prime focus camera capable of studying the widest-possible
wavelength range from the UV down to the NIR H-band.

The Large Binocular Camera (LBC, \cite{ragazzoni,pedik,pedik04,ragazzoni06})
is a wide FoV instrument at the prime focus of the twin 8.4 meter
Large Binocular Telescope (LBT).  The LBT uses two 8.4-meter diameter
honeycomb primary mirrors mounted side-by-side to produce a collecting
area of 110 square meters equivalent to an 11.8-meter circular
aperture.  A unique feature of the LBT is that the light from the two
primary mirrors can be combined optically in the center of the
telescope to produce phased array imaging of an extended field.
This requires minimal path length compensations, thus making
interferometry easier than in completely independent telescopes.

The requirement for an instrument such as LBC has been identified by several
high-profile scientific programs that call for an increase in FoV and
high-UV/IR sensitivity for deep imaging.

These attributes are essential to programs studying a large
FoV, to significant depth, over a wide spectral range, and
can only be provided by an imager mounted at the prime focus of an
8m-class telescope.

In Section 2 we provide a description of the two LBC cameras, while in
Section 3 we detail the technical performance of LBC-Blue during
commissioning observations in 2006. In Section 4 we analyze in
detail the case for an UV deep imaging survey in an extragalactic field,
and compare results with those obtained using different instruments
and telescopes. We present our conclusions in Section 5.


\section{The LBC camera}

The Large Binocular Camera (LBC) is a wide-field double
imager at the prime foci of the LBT.

\begin{figure}
\includegraphics[width=9cm]{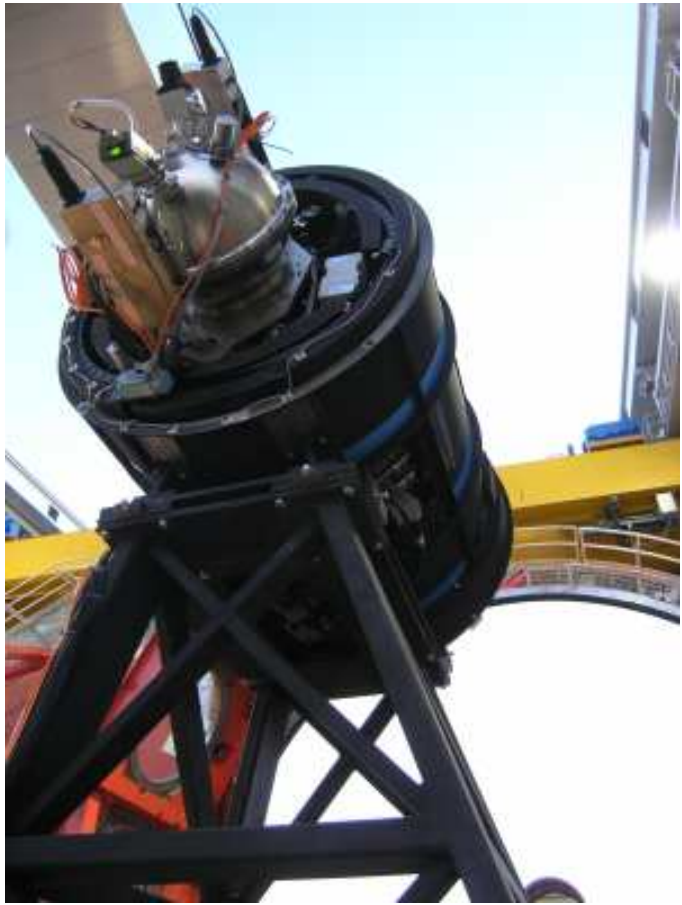}
\caption{
The LBC-Blue instrument installed at the prime focus of the first LBT
telescope unit.
}
\label{fig:lbcb}
\end{figure}

\begin{figure*}
\includegraphics[width=9cm]{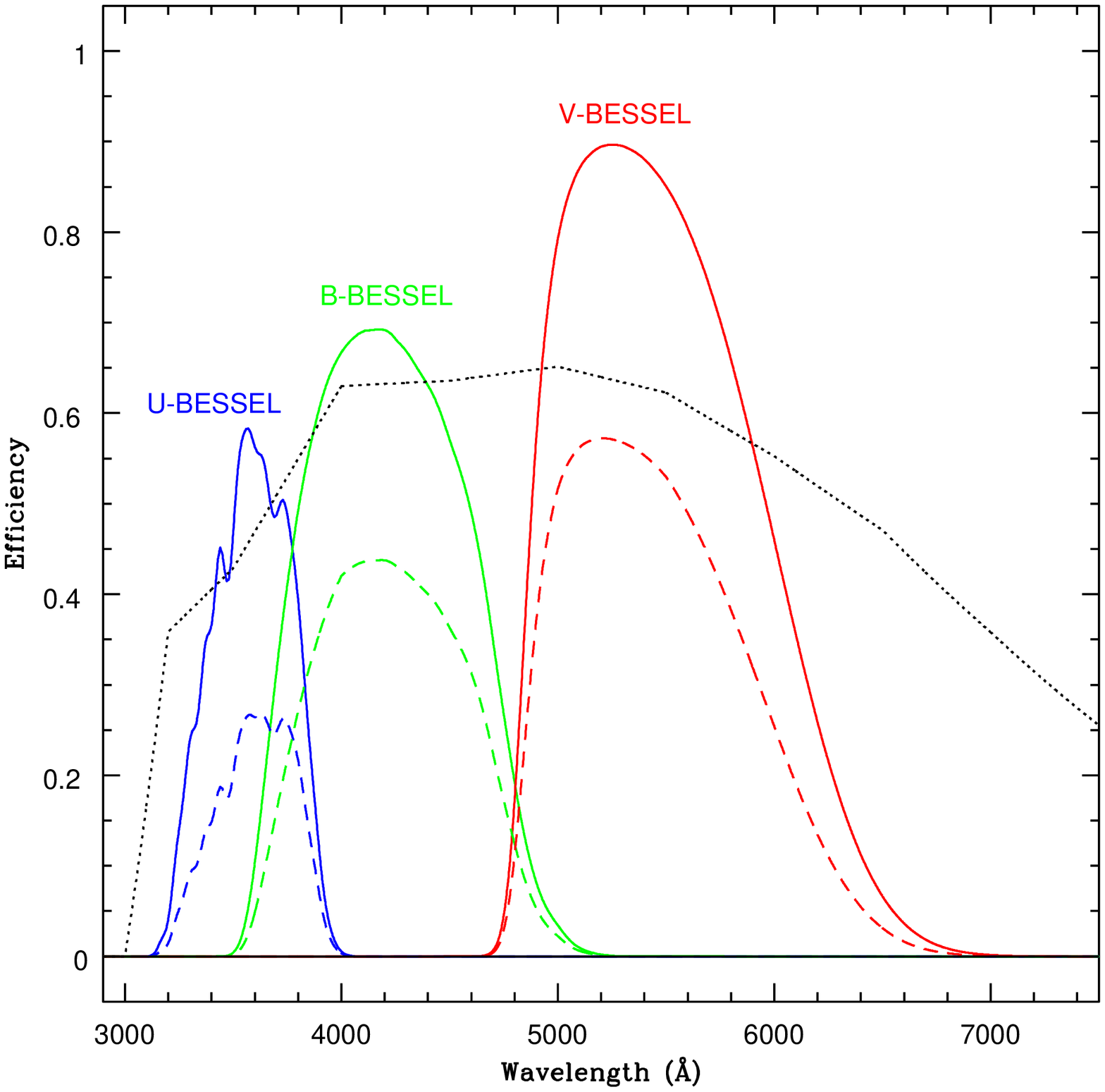}
\includegraphics[width=9cm]{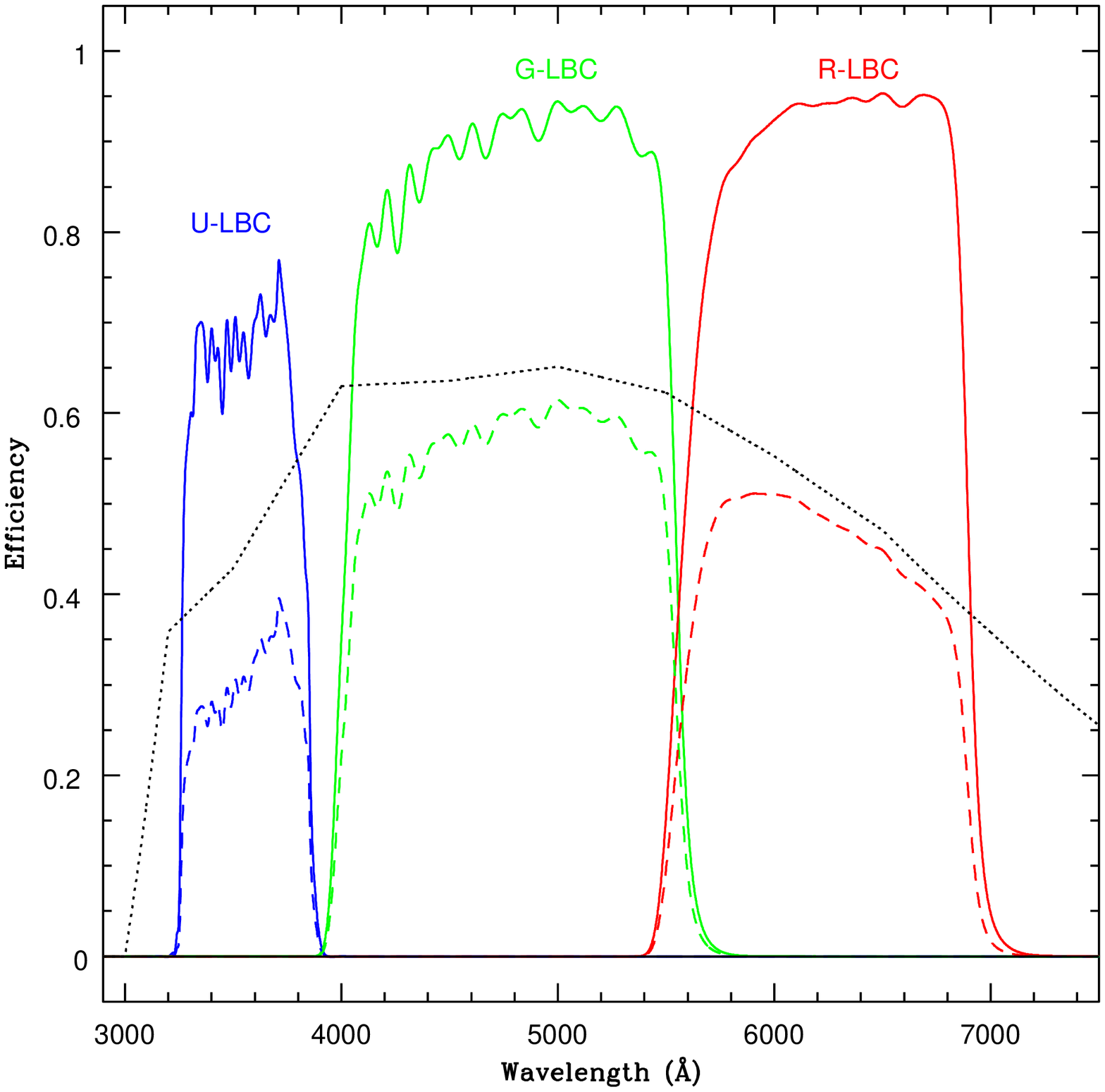}
\caption{
LBC-Blue filter set. Bessel U, B, and V in the left panel,
U-LBC, G-LBC, and R-LBC in the right panel.
Dashed curves are the filter response curves derived after convolution with 
the overall LBC-Blue efficiency (the combination of CCD efficiency,
mirror reflectivity, and optics transmission),
which is shown as a dotted line.
}
\label{fig:filterb}
\end{figure*}

The two channels are optimized for different wavelength ranges: the blue
channel (LBC-Blue) for the U, B, and V bands, and the red channel (LBC-Red)
for the V, R, I, and Z bands.  Fig.\ref{fig:lbcb} shows the
LBC-Blue instrument installed at the LBT, while Fig.\ref{fig:filterb}
describes the efficiency of the filter set available for LBC-Blue.
The fast focal ratio (F/1.45) allows efficient deep imaging
over a FoV of approximately 30 arcmin in diameter.
Because the mirrors of both channels are mounted on the same pointing
system, a given target can be observed simultaneously over a wide
wavelength range, improving the operation efficiency.
The fast prime focus
configuration requires an optical corrector to compensate the
aberrations introduced by the primary mirror.  The unique binocular
configuration of LBT allowed the optimization of both correctors
for different wavelength ranges.
This simplifies the instrument design by relaxing the requirements on
the achromaticity for each channel.

\subsection{LBC-Blue}

\begin{figure}
\includegraphics[width=9cm]{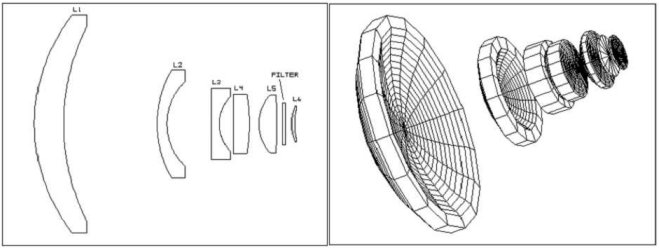}
\caption{
LBC-Blue. 2D layer and 3D model of the optical corrector.
}
\label{fig:lenses}
\end{figure}

The focal ratio of the primary mirror (F/1.14), the large telescope
diameter (8.4m), and parabolic shape of the mirror present particular
difficulties for the design of a prime focus corrector for the LBT.
The design of a prime focus corrector for the blue channel can be
described as a development of the design of \cite{wynne}, which
consists of three lenses to correct spherical aberration, coma,
and field curvature. In our
design the second and third lenses are split into two elements, and
with respect to Wynne's design an additional lens is present
that has a positive meniscus of almost no net power, which is the CCD
cryostat window. Due to the size of the primary mirror the
largest lens of the corrector has a considerable diameter (810 mm) and
weight (104 Kg). All the lenses are in fused silica, which ensures
high throughput in the targeted wavelength range.  The optical
surfaces are spherical or plane, except lens \#3, featuring an
aspherical surface on the concave side; this surface is actually
ellipsoidal and presents a departure from the best fit sphere of 0.7mm
at the edge (Fig.\ref{fig:lenses}).  Geometric distortion is not
considered as an aberration, since it may be corrected by
post-processing. Two filter wheels are placed between the last two
lenses.  The focal length of the optical corrector is 12180mm and the
final focal ratio is F/1.45. The total throughput is 84\%.  The
throughput considers the internal transmission of the SILICA lenses
and the coating efficiency; it is an average figure for the U and B
bands and does not consider the filter transmission.

\begin{figure}
\includegraphics[width=9truecm]{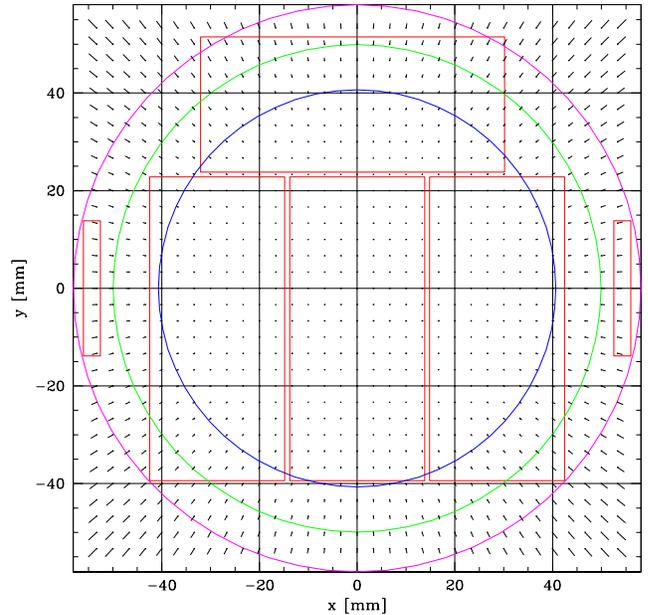}
\caption{
The optical distortion map of LBC-Blue. The inner, middle, and outer circles
mark the 1\%, 1.5\%, and 2\% distortion limits, respectively. The geometric
distribution of the four science chips and of the two technical arrays
is also shown.}
\label{fig:distortions}
\end{figure}

\begin{figure}
\includegraphics[width=9truecm]{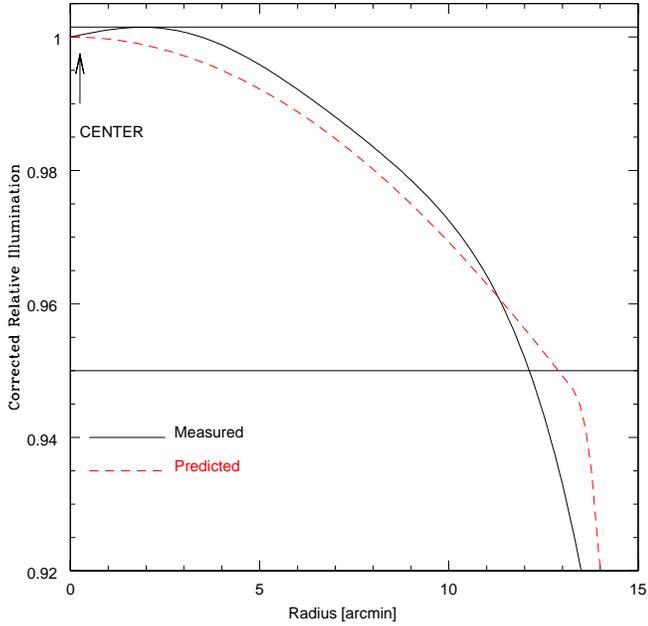}
\caption{
Flat-field illumination profile in the R band corrected for pixel
scale variation across the field. The expected profile from the
optical design is also shown for comparison.}
\label{fig:flatfield}
\end{figure}

The energy concentration of the instrumental PSF is very good: 80\% of
the energy is enclosed in a single CCD pixel (13.5$\mu$m in size or
0.2254 arcsec) both in the U and B bands which ensures good optical
performance even in the best seeing conditions (FWHM$\sim 0.4$
arcsec). Although the blue channel has been optimized for the U and B
bands, the performance is good also in the V and R bands, with
80\% of the energy within 2x2 pixels. The geometric distortion, of
pin-cushion type, is always below 1.75\% even at the edge of the field
(see Fig.\ref{fig:distortions}). The unvignetted FoV is 27 arcmin in
diameter, as shown in Fig.\ref{fig:flatfield}.

\begin{figure}
\includegraphics[width=9cm]{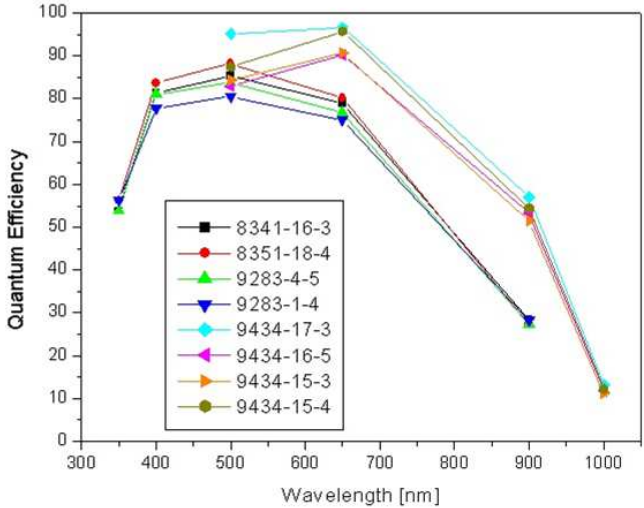}
\caption{Quantum efficiency of the science chips of the blue and red channel
(higher points at 900 nm).
}
\label{fig:chips}
\end{figure}

The mechanical design of the prime focus consists of two main parts,
the hub that mounts the fixed lenses and the derotator that holds the
filters wheels and the cryostat. Each one of the five fixed lenses is
kinematically mounted into an INVAR frame, which is then connected to
the steel hub through flexure elements to accommodate the differential
thermal expansion of the two materials. For the same reason, the two
main lenses, which are 810 mm and 400 mm in diameter, are mounted into
their INVAR frames by means of special RTV pads, that are tailored to
compensate for the differential thermal expansion of the glass and the
INVAR.

The derotator decouples the imager from the corrector lenses and hosts
two filter wheels, the shutter and the cryostat. The cryostat
mount enables all instrumental electronics to be mounted onto the
derotated structure.  Concerning the servo controls of the instrument,
each motor controller is addressed by the control PC as a network node
by means of a TCP-IP protocol.

Both cryostats were designed to cool down to 170K the
detector flange that holds the scientific array of each camera
composed of four E2V 42-90 chips and two more E2V detectors for
technical use. The cryostat is composed of three independent modules:
a stainless steel interface flange, a nitrogen vessel and a housing
made of aluminium. This configuration allows us to separate the
electrical part (detector flange, cables, etc) from the cryogenic
assembly, allowing an easy maintenance and upgrade independently
of the two parts.

From a mechanical point of view, the novel component of the blue channel
cryostat is the bimetallic and monolithic vessel. It was designed with a
spherical shape both to minimize the radiative thermal inlet and to
create a compact instrument. With this geometry we obtain both a
smooth cooling of the CCD baseplate and a good temperature stability,
the latter being independent of the position of the camera.  With a
10-liter fill of liquid nitrogen, the hold time of the cryostat is
approximately 48 hours.

Two types of E2V detectors have been mounted to ensure simultaneous
monitoring of the
scientific data acquisition and the control of the instrument: an
array of four E2V 42-90 (4608$\times$2048 pixels) chips cover the
corrected field with a sampling of about 0.2254 arcsec/pixel providing
a scientific image of $23.6\times25.3 arcmin^2$, while two E2V 42-10
of 256$\times$2048 px are used to acquire short exposure images for
guiding and wavefront control. The technical characteristics of the
42-90 are: QE$\gtrsim 80$\% at the peak, charge transfer efficiency
$>99.999$\%, read-out-noise $<5$ electrons at 1MHz, and surface roughness
$<7 \mu$m peak to valley. The final constraint is imposed by the instrument's
fast focal ratio (F/1.46), and is necessary to ensure optimal image quality
over the entire focal plane. The gaps between the
vertical chips are 1mm, which corresponds to 74 pixels or
equivalently 16.7 arcsec in the focal plane of LBC-Blue.  The gap
between the vertical chips and the horizontal chip is 1.03mm
(76 pixels, 17.2 arcsec).

There is a 5\% loss of energy in the blue channel at the edge of the
corrected field, while in the red channel the percentage of
vignetting is well below that for the corrected area.  The four science
chips are placed in an unconventional fashion, with the fourth
chip rotated 90$^{\circ}$ with respect to the others, to optimally
cover the corrected FoV (see Fig.\ref{fig:distortions}).

The CCD controller selected for the LBC camera was designed and
produced by the Italian firm Skytech in collaboration with the LBC
team. The core of the system is a programmable Xilinx FPGA used to
accomplish several different tasks.  The whole system is compact and
uses only two half eurocard boards to be better hosted at the prime
focus application of the LBC. A good noise performance of 11$e^-$ at 500
Kpix/s/ch is achieved despite the lack of a video preamplifier.
In this configuration, the total readout time of the CCDs is 27
seconds.

The LBC shutter adopts dual blade mechanics to ensure a uniform
exposure on the overall field also at short exposure times (0.1
sec).  The accuracy ($\sim 2/1000$ sec) has been measured by
laboratory tests using a laser trap.

Two filter wheels with five holes are available for each channel. At
present U, B, and V Bessel filters as well as custom U, G, and R
filters are available for the blue channel. Their spectral shapes
convolved with the LBC efficiency are shown in Fig.\ref{fig:filterb}.
The G filter of LBC in practice is equivalent to a standard Gunn-g
filter, while the U and R filters have been custom-made for the LBC-Blue
instrument. The corresponding physical size of the filters is
155mm of diameter while the shutter used to cover the entire FoV of
LBC-Blue is 470mm$\times$186mm (\cite{speziali}).

The operation of the camera is handled by a graphical user interface
and all the raw, calibration and telemetry data obtained so far are
publicly available for the LBT partners in the LBC archive.

\subsection{LBC-Red}

The red channel corrector is optimized for the wavelength range
including the V, R, I, and Z bands, with a possible extension to the
near infrared, up to 1.8 $\mu$m (J and H bands).

Although the wavelength range of interest is approximately two
times larger than for the blue channel,

the design of the red channel corrector was easier because of a
smaller change of refractive index with wavelength, toward
the red part of the visible spectrum.
Two different glass types were considered, silica and BK7. It should be
stressed that neither silica nor BK7 ensure optimal transmission in
the near infrared; on the other hand the main use of the instrument
will be in the wavelength range from V to Z band and the extension to
J and H should be intended as an additional facility. After a careful
evaluation, BK7 was chosen, due to its better optical
performance and lower cost.
The instrumental sensitivity in the near infrared depends on the
ability to change the final lens (L6) when replacing the
detector for the infrared cryostat.

The red channel design is similar to that of the blue channel, with 6 lenses
(5 spherical and 1 aspherical) and a plane filter. The focal length,
and hence the plate scale, are almost equivalent to the the blue channel
design. We have in addition tried to ensure that geometric distortion is
similar for both channels to simplify the development of data
reduction tools. The energy concentration at the wavelengths of
interest is always well within the goal of 80\% of the input energy to a
single CCD pixel. The total throughput is 82\%.

The red channel mechanical design is similar to that of the blue
channel, considering the size of the lenses, their distances and
especially the focal plane region and filter wheels.
The only major disparity between the performance of the two channels
is due to the different thermal behaviours of BK7 and SILICA.

The red channel is equipped with 4 high-resistivity, deep-depletion
42-90 E2V detectors, which are optimized for high efficiency at longer
wavelengths.  The QEs of the four chips are shown in
Fig.\ref{fig:chips} and the technical characteristics are similar to
the chips selected for the blue channel.

The red camera was commissioned at the end of 2007 and the
binocular configuration of LBT will be available at the beginning of 2008.
Further details about the instrument are available on the LBC web site at
{\sl http://lbc.oa-roma.inaf.it/}.


\section{The commissioning run of LBC-Blue}

The LBC-Blue camera was the first instrument installed at the LBT, and
for this reason was also used during the commissioning of the first LBT
telescope.

During the commissioning phase completed in
October-December 2006, a large number of astrometric and photometric fields
were observed to be able to characterize the instrument. A detailed
description of observations is provided in the next section. Some nights were
dedicated to observing scientific targets to be able to test the performance of
the instrument.

\begin{figure}
\includegraphics[width=9cm]{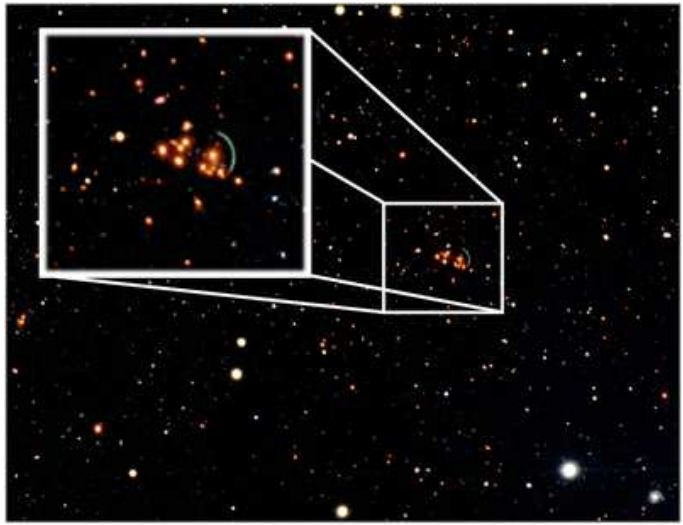}
\caption{
The galaxy cluster CL2244-02 and its gravitational arc as seen in the
central chip of LBC-Blue (only $\sim$10\% of the entire FoV is shown
here).  The FWHM is particularly good, 0.45 arcsec in the U band and
0.55 in the B and V bands.
}
\label{fig:cl2244}
\end{figure}

The scientific targets were selected to assess the ability of LBC
to address a wide range of open science questions in the near future.
We observed, for example, the nearby
galaxy cluster CL2244-02, which acts on a distant galaxy producing a
spectacular gravitational arc observed by LBC, during a night of 0.45
arcsec seeing in the U band (see Fig.\ref{fig:cl2244}).
Our imaging data demonstrates that the LBC can provide high-quality
data to study strong, gravitational lensing, provided that an
efficient service-observing program, to assess current seeing
conditions, is in place. Several star clusters were observed, NGC7789,
NGC2419, and M67, to analyze the technical performance related for example to
the variation of the PSF profile across the overall LBC FoV, and to the
analysis of astrometric distortions due to the complex optical
correctors described in Section 2.  Extragalactic targets, such as the
galaxy cluster Abell576, the Subaru XMM Deep Survey (SXDS) and the
quasar Q0933+28 field were observed to derive the magnitude limits in
deep imaging surveys, and to add deep UV-B images to the multicolour
information already available in these fields for the analysis of the
evolutionary properties of faint, distant galaxies.

\subsection{Technical performance of LBC-Blue}

The presence of a pre-scan and an over-scan in each image allows a
correct subtraction of the CCD bias signal which is stable in time.
Dark current is negligible because exposure times of a single image
are typically smaller than 15 minutes.  The read-out noise and e-/ADU
conversion factor (gain) have been measured by applying the ``variance
method'' to flat field sequences and were found to be in agreement with our
laboratory measurements. Chip to chip small variations were measured and
are summarized in Table \ref{tab:rongain}.

The camera provides a linearity residual error smaller than 1\% over the
whole 16-bit dynamic range (120000 e$^-$). The detector full-well limit
before blooming is greater than 150000 e$^-$.

In most imagers based on CCD mosaics, electronics ghosts due to the
video channel's cross-talk are often present and are removed using
software algorithms specifically developed for this purpose (see for example
{\em http://lbc.oa-roma.inaf.it/commissioning/xtalk.html})
especially when bright saturated sources
are present in the field of view. For LBC the cross talk coefficients
are always of the order of $3\times 10^{-5}$.

\begin{table}
\caption{Read-out noise and gain factors for the four chips of LBC-Blue.}
\scriptsize
\begin{tabular}{ccccc}
\hline
\hline
\textbf{Parameter} & \textbf{Chip \#1} & \textbf{Chip \#2} &
\textbf{Chip \#3} & \textbf{Chip \#4} \\
\hline
\textbf{R.O.N. e$^-$} & 11.4 & 11.6 & 11.6 & 11.2 \\
\textbf{e$^{-}$/ADU} & 1.96 & 2.09 & 2.06 & 1.98 \\
\hline
\hline
\end{tabular}
\label{tab:rongain}
\end{table} 

Flat-fielding was completed using a combination of twilight sky and
night sky data. In Fig.\ref{fig:flatfield}, we show our flat-field
illumination correction as a function of radius from the center of the
field-of-view, calculated using observations acquired in April
2007. The dome-shaped profile is the result of intrinsic illumination
pattern, with residual scattered light affecting the central region of
the FoV and vignetting affecting the outer regions of the FoV.  All
the curves are normalized to unity at the center. In the figure, the
contribution of the sky ghost to the radial profile can be seen at a
level of 0.15\%, close to the geometrical FoV center (pixel x=1024 y=2919 of
chip no. 2). The radial profile, predicted by our original design, is
shown for comparison.  Differences between the profiles are below a
level of 1\% out to 13 arcminutes from the center of FoV.

Being a prime focus camera, distortions are expected to be
significant. The original design predicted that the effects of distortions
would be fully-corrected out to about 10 arcmin from the FoV center.
Distortions of light lead to two main problems, spatial variations in
the pixel scale and deformations of the image PSF.
The first problem has
been computed from simulations and tested with images of moderately,
crowded fields.  It is corrected through an astrometric solution
process estimated from the images.  The second problem is partially
recovered by resampling the images and using a variable PSF model, or
using sufficiently large photometric apertures.
We applied both methods, in each case by measuring the flux of
standard stars, and found the results of both to be in close
agreement.

The main optical distortion map is reproduced in
Fig.\ref{fig:distortions}.

The astrometric solution was calculated in a three-step process, developed
using the software package
\textit{AstromC} written by M. Radovich: this package is a porting
to C++ of the Astrometrix software described e.g. in \cite{radovich}.
The final astrometric solution is similar to the theoretical
pre-solution derived from the original optical design.  Second-order
corrections vary from frame to frame because of different elevation,
filter or position angle.  These variations are however very small.
Filter-to-filter variation is of the order of 0.01\%, corresponding to
about 1 pixel at the edge of the FoV.  The pixel scale at the center
is 0.2275$\arcsec \pm$ 0.0001 and the median value is 0.2254$\arcsec \pm$
0.0001 with filter-to-filter variation
affecting the fourth decimal digit.  The corrected individual frames
can be resampled to a constant pixel scale and stacked to a final
image mosaic.

\begin{figure}
\includegraphics[width=9cm]{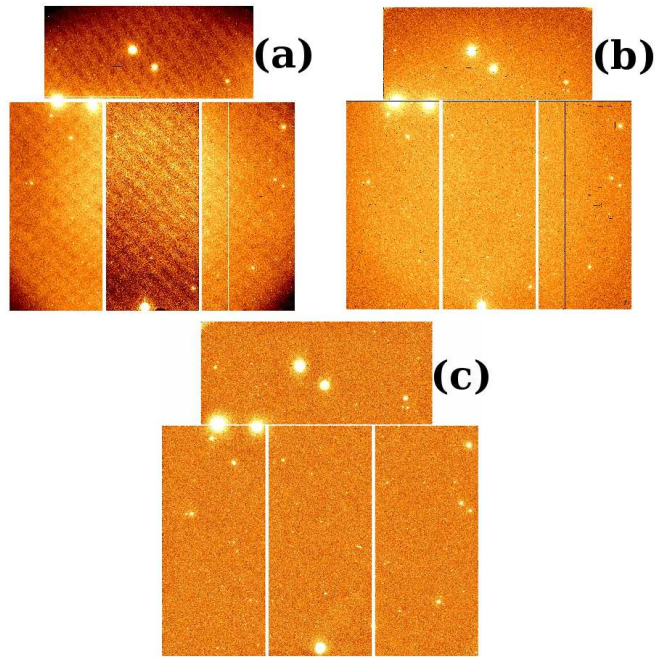}
\caption{(a) Raw LBC-Blue image. (b) The same image after
flat-fielding correction. (c) The same image after re-sampling to a constant
pixel scale.}
\label{fig:fov}
\end{figure}

In Fig.\ref{fig:fov} we show an example of the mosaic LBC-Blue field
of view in the U-Bessel filter, after applying various steps of the
reduction procedure. The raw image in Fig.\ref{fig:fov} (a) shows both
the effects due to vignetting and sky concentration caused by the
geometrical distortion. In Fig.\ref{fig:fov} (b) the same mosaic image
is shown after cross-talk correction, bias subtraction and
flat-fielding, which removes the vignetting effect.  In
Fig.\ref{fig:fov} (c) the science mosaic is shown after removal of the
geometrical distortion by re-sampling to a constant pixel scale. The
quality of the PSF over the entire FoV of LBC-Blue does not depend on
the radial distance from the optical center.  This is also true for
the ellipticity of stars in the field, which is always below 0.05
with a median value of 0.02.  Thus, the uncertainties in telescope
guiding and the optical distortion do not significantly affect the
quality of the image at large distances from the center.

We have also analyzed images taken during the commissioning and
science demonstration time (SDT), to quantitatively estimate the
presence and magnitude of ghosts due to bright stars.  The Bessel U,
B, V, and custom G and R filters do not show measurable ghosts, in
agreement with our expectations.  Indeed, according to simulations,
the ghost due to a bright star falls at the center of the bright star
itself and it is twelve magnitudes fainter. In the interference U-LBC
filter, the primary ghosts due to bright stars are more evident.  The
primary ghost of a UV bright star has the shape of a diffuse
circle. To estimate the amount of flux in the primary ghost, we first
built a model for the PSF using stars with peak intensity less than
20000 ADU and then used this model PSF to fit the non-saturated wings
of bright stars.  After PSF subtraction, two components remain: an
intense circular ring centered on the bright star (diameter=75 pixels),
and a diffuse large component (diameter=200 pixels) of smaller
intensity, shifted with respect to the bright star, in radial direction
from the center of the FoV.  We
have computed the total intensity of the inner bright component and of
the diffuse, larger one finding a value of 2.8$\pm$0.7 per cent.  We
do not find evidence of dependence of the ghost intensity on the
distance from the center field. The ghost intensity is independent of
the position angle of the camera or elevation of the telescope. No
secondary ghosts are measurable on the science data.

Calibration equations  have been  obtained for  two photometric nights
during the commissioning run of November 2006 in the Bessel U,
B, and V filters. The equations adopted were of the kind:
\begin{equation}
V  = v + z.p.  + c.t. \times  (b-v) + k_V  \times X_V \,
\end{equation}
where \textit{z.p.}  is the photometric zero point, \textit{c.t.} is
the colour term, $k_i$ is the atmospheric extinction coefficient in the
$i$-th band and $X_i$, the airmass in the same band. We calibrated V
and B versus the colour $b-v$, while filter U has been calibrated
versus $u-b$.  Photometric standards have been obtained from the
\cite{landolt92} catalog, namely from the fields SA98 and SA113. These
fields also contain a large number of additional standards measured by
P.B. Stetson and available from the CADC web site.  A supplementary
catalog from \cite{galadi00} has been used as well to increase the
sample of U standards (Stetson standards have only B, V, R, and I band
magnitudes) and to have a standard field close to the zenith for a
better fit of the atmospheric extinction.  The fits obtained are
remarkably good and their internal errors very small.  The photometric
accuracy in the overall field is of the order of 0.01 mags.  The
values of the coefficients, and their uncertainties are
reported in Table \ref{tab:photcalib}.  For the G and R filters, only
zero points have been obtained so far using SDSS secondary standard
fields, while the U-LBC filter was not available during the
commissioning phase in 2006.

\begin{table*}[ht]
\begin{center}
\caption{Photometric calibration coefficients}
\begin{tabular}{ccccc}
\hline
\hline
\textbf{Filter} & \textbf{Zero Point} & \textbf{Colour Term} &
\textbf{Extinction} & \textbf{corrAB} \\
\hline
U-BESSEL & 26.23 $\pm$ 0.03 & 0.036 $\pm$ 0.009 (u-b) & -0.48 $\pm$ 0.02 & 0.87 \\
B-BESSEL & 27.93 $\pm$ 0.02 & -0.123 $\pm$ 0.003 (b-v) & -0.22 $\pm$ 0.01 & -0.07 \\
V-BESSEL & 28.13 $\pm$ 0.01 & 0.021 $\pm$ 0.005 (b-v) & -0.15 $\pm$ 0.02 & 0.01 \\
G-LBC & 28.59 $\pm$ 0.02 &  & & -0.11 \\
R-LBC & 27.63 $\pm$ 0.02 &  & & 0.13 \\
\hline
\hline
\end{tabular}
\label{tab:photcalib}
\\
The zero points are in the Vega photometric system, and the AB magnitude
can be derived using the relation $mag(AB)=mag(Vega)+corrAB$.
The U, B, and V filters are the Bessel ones, while G and R filters are
peculiar of LBC-Blue. Only zero points are available
for the G and R filters. The U-LBC filter was not available during the
commissioning phase in 2006.
\end{center}
\end{table*}


\subsection{A test for deep imaging in the UV: the Q0933+28 field}

To test the capabilities of the LBC-Blue channel in the field of deep
extragalactic surveys over large areas of the sky, we observed a field
centered on the bright QSO Q0933+28 at z=3.42 (\cite{veron}).
This field has been observed by \cite{steidel} in the UGR filter set
and was the target of an intensive spectroscopic campaign by Steidel
and collaborators, which generated hundreds of galaxy spectra down
to $R=25.5$ AB mag in the redshift interval $1\le z\le 4$.
This field is therefore ideally-suited as a target to
test the performance of LBC in extragalactic astronomy.
In particular a deep U-BESSEL image has been obtained in this field.
The details of the observations are provided in Table \ref{tabq0933}.

Raw LBC images were reduced using the LBC Pipeline, a collection of C
and Python scripts optimized for LBC data analysis. The software
performs cross-talk correction, bias subtraction (line by line,
fitting the pre-scan and the over-scan), flat-field
normalization. Then, we applied the astrometric solution given by
AstromC to the provided frames and stacked them into a single mosaic
using the Swarp package\footnote{\sl http://terapix.iap.fr/}.  The
astrometric procedure uses positions and fluxes from overlapping
sources in different exposures to simultaneously optimize the internal
astrometric accuracy and derive a relative photometric calibration of
the stacking.  Then absolute calibrations were obtained using
photometric standard fields (Landolt and/or Stetson) observed in the
same night. To calibrate the G and R filter zero points
we used the photometry in the Stone fields (\cite{stone}).
Part of the LBC field during commissioning was affected by scattered
light, for this reason the total area of the Q0933+28 field was
limited to 478.2 arcmin$^2$.

\subsection{Deep U band galaxy counts}

To measure the efficiency of LBC-Blue for deep photometry in the UV
bands over a large FoV, we acquired 3 hours of observations of the
Q0933+28 field in average seeing conditions ($FWHM\sim$ 1 arcsec).
We then used SExtractor (\cite{sex}) to derive a photometric catalog,
and computed U-band galaxy number counts.
For objects with area greater than that
corresponding to a circular aperture of radius equal to the FWHM,
we used the isophotal magnitudes provided by SExtractor.
For smaller sources we computed magnitudes in circular apertures with
diameter equal to 2 times the FWHM. This allows us to avoid the well-known
underestimate of the flux of faint galaxies provided by the isophotal
method. To isolate the few stars from the numerous faint
galaxies in this field, we relied on the class\_star classifier provided
by SExtractor.

\begin{figure}
\includegraphics[width=9cm]{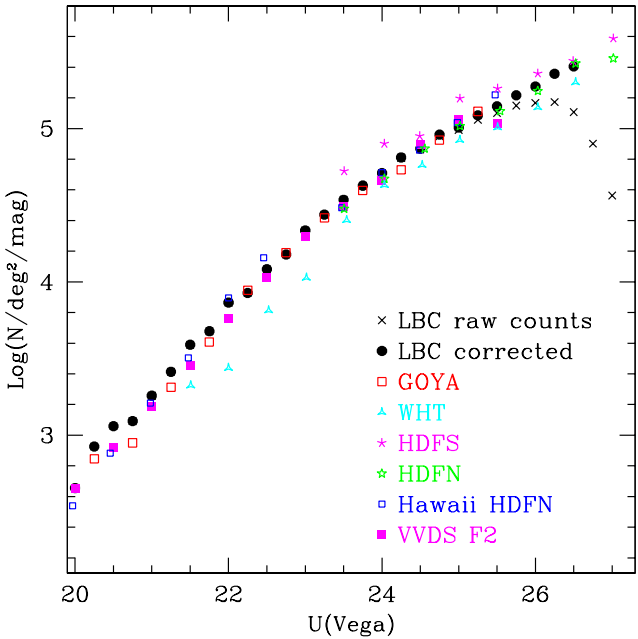}
\caption{
Number counts of galaxies in the U-BESSEL band for the Q0933+28 LBC
field. Magnitudes are in the Vega system. We compare our counts with
shallow surveys of similar area (GOYA, Hawaii HDFN, VVDS F2), and with
deeper pencil beam surveys (WHT, HDFN, HDFS).
}
\label{fig:lognsu}
\end{figure}

Raw counts are shown in Fig.\ref{fig:lognsu} where a clear decrease
is apparent for $U(Vega)>26.4$.
Thus an estimate of the completeness level should be performed in
order to evaluate the amount of correction to the raw counts at the
faint limits.  This has been evaluated including in the real image
1000 simulated galaxies per magnitude bin in the magnitude interval
$U(Vega)=24-27$ using the standard ''artdata'' package in IRAF.  For
simplicity disk galaxies are included with convolved sizes typical of
real galaxies in the magnitude interval $U(Vega)=24-25$, i.e. with
$FWHM\sim 2.5$ arcsec.  The resulting 50\% completeness level is measured
at $U(Vega)=26.5$. The corrected counts are shown in the same
Fig.\ref{fig:lognsu}.  Given the wide magnitude interval from
$U(Vega)=19$ to $U(Vega)=26.5$ available in the present survey, the
shape of the counts can be derived from a single survey in
self-consistent way, possibly avoiding offsets due to systematics in
the photometric analysis.  A clear bending is apparent at $U(Vega)>
23.5$. To quantify the effect we fitted the shape of the counts in the
above magnitude interval with a double power-law. The slope changes from
$0.62\pm 0.1$ to $0.22\pm 0.04$ for magnitudes fainter than $U_{break}=23.2$.
The uncertainty in the break magnitude is however large, $\sim 0.8$.

In Fig.\ref{fig:lognsu} we compare our number counts with those
derived by shallow surveys of similar area (GOYA by \cite{goya};
Hawaii HDFN by \cite{capak04}; VVDS-F2 by \cite{radovich}), and with
deeper pencil beam surveys (WHT, HDFN, and HDFS by \cite{wht}). In
particular, the WHT galaxy counts (\cite{wht}) are based on a 34h
exposure time image reaching $U(Vega)=26.8$ but at the much lower
3$\sigma$ level in the photometric noise and in an area of $\sim$50
arcmin$^2$, while the GOYA survey at the INT telescope is complete at
50\% level at $U(Vega)=24.8$. These counts are shown together with the
two pencil beam surveys in the Hubble Deep Fields (Metcalfe et
al. 2001).

The agreement with the GOYA survey (900 sq. arcmin.) is remarkable,
and suggests that once large areas of the sky are investigated, the
effects of cosmic variance are slightly reduced.  Deep pencil beam
surveys (HDFN, HDFS) can go about 1 magnitude deeper than our
present magnitude limit but require much longer exposure times.  The
present UV counts obtained during the commissioning of LBC-Blue are
the deepest obtained so far from ground-based observations in large
sky area that are not affected by cosmic variance.  Deeper
observations are expected with LBC-Blue in fields with larger exposure
times and with more efficient UV filters.

It is interesting at this point to compare the LBC-Blue performance in
particular with that of MegaCam at CFHT, because Suprimecam at Subaru is
not efficient in the UV. LBC is of course 4-4.5 times faster than
MegaCam at CFHT in the UV-B bands but the field of view is about 1/6
deg$^2$.  Thus LBC is optimized for very deep images on relatively
smaller areas. We note however that the high LBC UV efficiency allows
the use of UV filters centered at shorter wavelengths (355 nm) with respect
to the MegaCam one (375 nm) providing UV magnitudes very similar to
the standard Bessel system. Moreover, considering its final binocular
configuration the LBC camera will double its global efficiency in
multicolour imaging.

\begin{table}
\caption[]{Observations of the Q0933+28 field with LBC-Blue}
\begin{tabular}{ccccc}
\hline
\hline
Filter & maglim(10$\sigma$) & maglim(1$\sigma$) & seeing &
texp \\
\hline
 & Vega mag & Vega mag & arcsec & hours \\
\hline
U-BESSEL & 25.13 & 27.73 & 1.06 & 3.00\\
G-LBC & 26.01 & 28.51 & 0.83 & 0.45\\
R-LBC & 24.97 & 27.47 & 1.22 & 0.50\\
\hline
\hline
\end{tabular}
\label{tabq0933}
\\
The magnitude limits are computed in a circular aperture two
times the FWHM in each band. 
\end{table}

\subsection{UV dropout galaxies at high redshifts}

The field we have selected for the commissioning test of the blue
channel is one of Steidel's fields used for the search of Lyman break
galaxies at $z\sim 3$. For this field original U, G, and R images, a
multicolour UGR catalog and spectroscopic information are made
publicly available by Steidel and collaborators and are used for
comparison with the LBC images.

\begin{figure*}
\includegraphics[width=9cm]{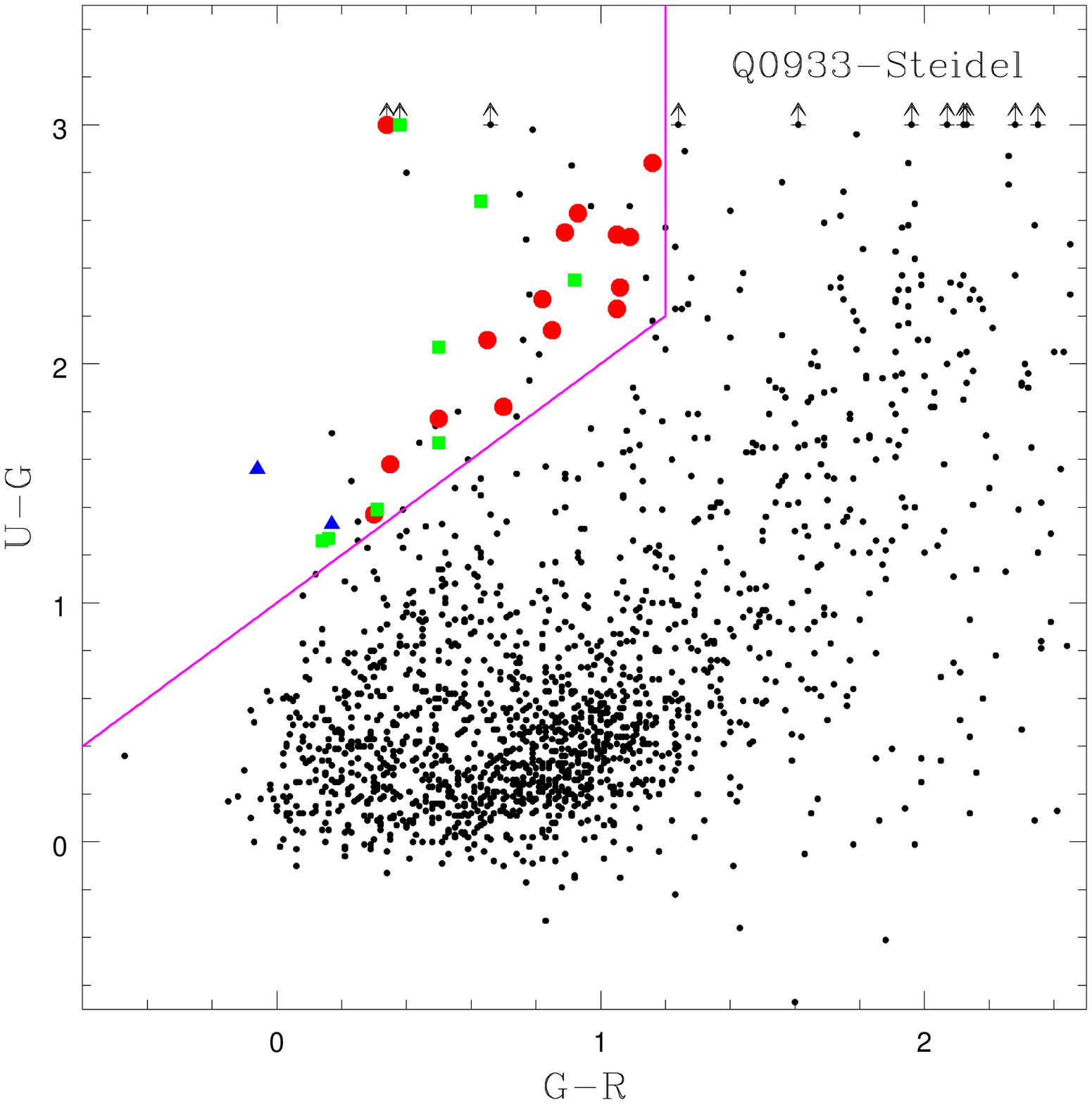} 
\includegraphics[width=9cm]{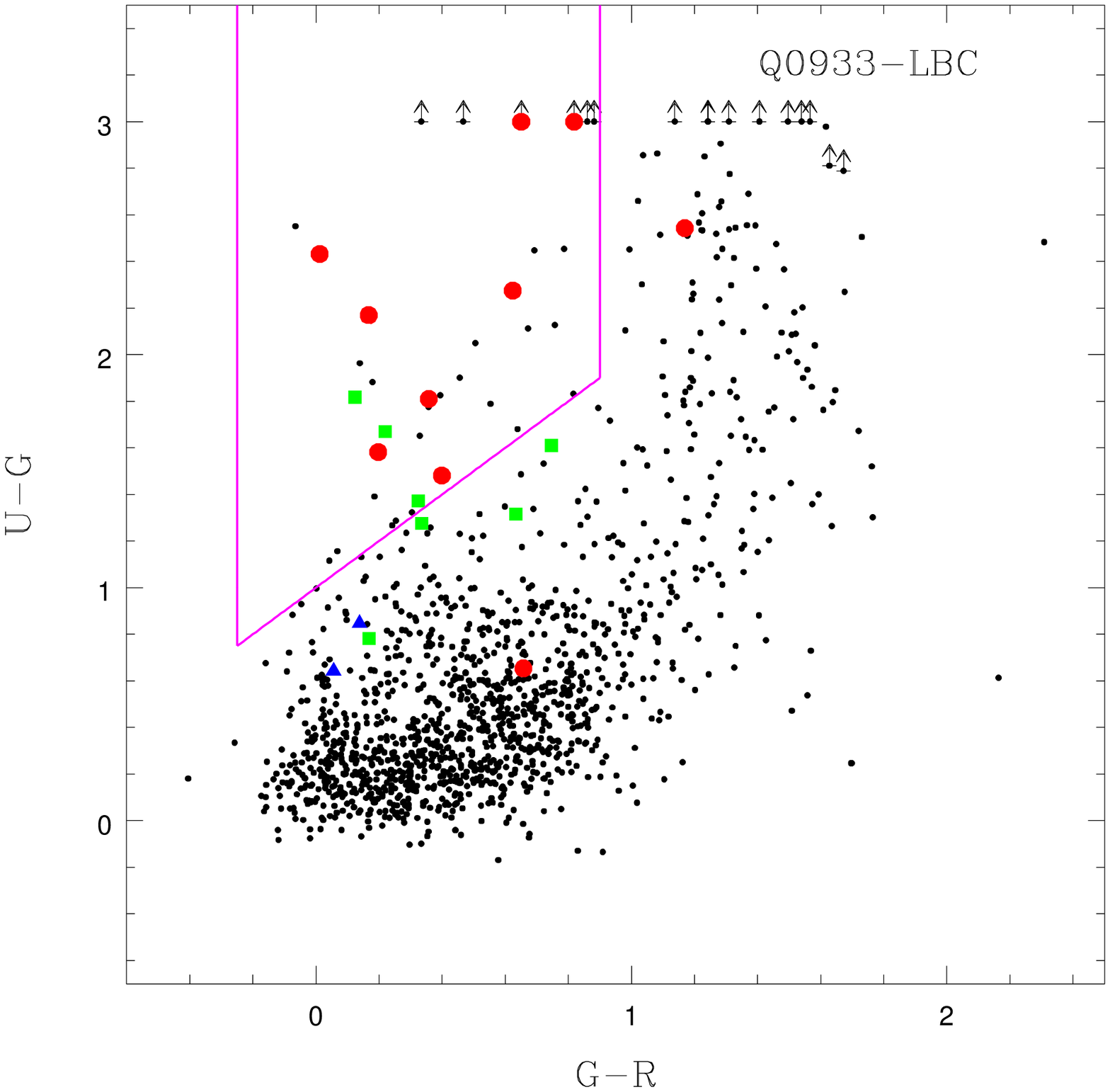}
\caption{Left panel:
selection of LBGs in the field by Steidel in the $U-G$, $G-R$ colour
diagram. Circles, squares, triangles are star-forming galaxies at $R\leq 24.5$
spectroscopically confirmed at $2.8<z<3.7$, $2.2<z<2.8$, and
$1.4<z<2.2$, respectively. Right panel: same as in the left panel but
using the colour catalog from the
LBC images. Note that some galaxies at $z<2.8$ are now
correctly out of the LBGs region.}
\label{fig:ugrstzspec}
\end{figure*}

It is well known that an efficient method to select unobscured or
modestly obscured star forming galaxies at high redshifts is the Lyman
break technique, that is effective in detecting high redshift Lyman
break galaxies (LBGs) at $2.8\le z\le 3.7$. This selection, adopted by
Steidel and collaborators (e.g. \cite{steidel99}; \cite{adelberger04})
is based on the UGR filter set and exploits the strong absorption
present in the UV band of these galaxies (UV dropout) caused by the
redshifted Lyman continuum absorption produced by the interstellar
neutral hydrogen of the same galaxies.  Extensive spectroscopic follow
up showed this multicolour selection to be highly effective
(see e.g. \cite{adelberger04}).

We have applied the same method with our UGR filter set of the blue
channel to test by means of a deep U band image the advantage of an
efficient UV imager at an 8m class telescope.

We have used the R image of the Q0933+28 field to obtain a catalog of
galaxies whose magnitude limits at different $\sigma$ levels are
reported in Table \ref{tabq0933}. We have used SExtractor (\cite{sex})
in dual-image mode to derive the photometry in the other filters for
sources detected in the R band, which was used as the detection image.
We convolved images in each band with a Gaussian kernel so that the
equivalent seeing of each image was identical and equal to that
measured in the image taken in the worst seeing conditions,
i.e. R-band observations. In this way the stellar FWHM
of the images are the same and the colours can be computed in the same
object area.

The U-BESSEL and G filters of LBC-Blue are similar to the U and G
filters used by Steidel and collaborators. The R-LBC filter however
has a significantly different throughput than the R-band filter used
by Steidel et al. (2003), being peaked at shorter wavelength due
to LBC-Blue throughput decline above wavelengths of $\lambda=$7000\AA.
Comparing our
multiband photometry with the publicly-available catalog of the
Q0933+28 field by \cite{steidel} and on the basis of synthetic colours
predictions by galaxy spectral synthesis models, we derived the colour
equation to translate Steidel's colours into the LBC photometric
system, $(G-R)_{LBC}=0.7(G-R)_{Steidel}+0.05$.

We have modified the colour selection criteria of \cite{adelberger04}
taking into account the difference in the G-R colour and obtaining
the following colour selections:
$-0.25 \le G-R \le 0.9$ and $U-G \ge (G-R)+1.0$ for Lyman break galaxies
in the redshift interval $2.8<z<3.5$.

In Fig.\ref{fig:ugrstzspec} (left panel) we plot the U-G versus G-R
colour for all galaxies in the Q0933+28 field using the original
photometry by Steidel and highlight the selection criteria for LBGs.
We indicate the spectroscopically confirmed LBG galaxies
by Steidel's team, with 10 out of 24 galaxies being measured to be
at lower redshifts
($z<2.8$) although being in the colour region expected for LBGs. Only
two out of these ten galaxies are found at redshift $\le 2.2$.

In Fig.\ref{fig:ugrstzspec} (right panel) we show the same plot
derived from the U, G, and R LBC images. In this comparison we restrict
the analysis to the original Steidel's area of $\sim 9\times 9$
arcmin$^2$.
We show with triangles, squares, and circles the galaxies in
the redshift ranges analyzed by Steidel and collaborators:
$1.4<z<2.2$, $2.2<z<2.8$ and $2.8<z<3.5$, respectively. In this case
most of the LBGs candidates confirmed to be at lower spectroscopic
redshifts are in general bluer, i.e. brighter in UV and in most cases
lie consistently outside (or nearby, given the photometric noise) the
LBGs colour region.  Thus, even limiting the galaxy catalog to
relatively bright objects with $R<24.5$ the robustness of the UV
dropout colour selection technique of LBG galaxies at $z\sim 3$
increases when very deep UV images as obtained by LBC are used.

An attempt in this direction was recently performed by \cite{sawicki05}
who used very deep UGRI images obtained at Keck to produce a fainter
sample of Lyman break galaxies at redshifts $z=2,3,4$. However the
greater UV sensitivity and larger field of view of LBC makes this
instrument ideal to look for high redshift galaxies especially in the
context of the study of the large scale structures at high redshift.


\section{Conclusions}

In this paper we have described the first instrument at the LBT
telescope, the prime focus large binocular camera (LBC). The
instrument has a binocular configuration with two channels, the blue
channel with a good overall efficiency in the UV band and the red
channel with good efficiency in the V-z bands.

We have also shown the technical characteristics of the blue channel
derived from the commissioning data of LBC-Blue.

\begin{itemize}
\item
The corrected optical field is $\sim$ 30 arcmin of diameter with
$<5$\% loss of energy within $\sim 25$ arcmin. The total throughput of
the optical corrector is 84\%.
\item
The optical distortion is always $<1.75$\% even at the edge of the
field and it is removed with a specific SW package.
\item
The optical quality ensures an energy concentration of 80\% within a
pixel of $\simeq 0.2254$ arcsec in the overall corrected field.  The
active control of the optical quality can provide images as sharp as
FWHM=0.5 arcsec even in the UV band.
\item
Ghost images produced by the whole optical system are always
negligible when glass filters are used (Bessel U,B,V). The total
intensity of the primary brightest ghost is $\sim 2.8$\% when the wide
interference UV filter is used. The primary ghost is centered on the
original source position. Ghost images produced by the electronics
cross-talk are as small as $3\times 10^{ -5}$ and can be easily
removed during data reduction.
\end{itemize}

We have also described some scientific observations planned for the
commissioning to assess the performance of LBC-Blue.

\begin{itemize}  
\item
We have obtained very deep UV galaxy counts in a deep pointing with a
total exposure time of 3h reaching U(Vega)=26.5, after correction for
incompleteness.  The wide magnitude interval ($U(Vega)=19-26.5$) in
our galaxy counts allowed a direct evaluation of the shape of the UV
counts which shows a break in slope at about $U(Vega)=23.2$ with a change in
slope from 0.62 to 0.22 at the faint end.
\item
The same LBC area includes a quasar field where extensive study of
Lyman break galaxies at redshift $z\sim 3$ is available. We have
reproduced with our UGR filter set the well known multicolour
selection of LBGs showing that the robustness of the UV dropout method
for the selection of star-forming galaxies at $z\sim 3$ increases when
very deep UV images can be used as obtained by LBC at an 8m class
telescope like LBT.
\end{itemize}


\begin{acknowledgements}
Observations have been carried out using the Large Binocular Telescope
at Mt. Graham, Arizona, under the Commissioning of the Large Binocular Blue
Camera.
The LBT is an international collaboration among institutions in the
United States, Italy and Germany. LBT Corporation partners are: The
University of Arizona on behalf of the Arizona university system;
Istituto Nazionale di Astrofisica, Italy; LBT
Beteiligungsgesellschaft, Germany, representing the Max-Planck
Society, the Astrophysical Institute Potsdam, and Heidelberg
University; The Ohio State University, and The Research Corporation,
on behalf of The University of Notre Dame, University of Minnesota and
University of Virginia.
We thank C.C. Steidel and collaborators for the public availability of
their images and spectroscopic redshifts of the Q0933+28 field.
We thank the anonymous referee for useful comments which improve the
clarity of the paper.
\end{acknowledgements}


\begin{thebibliography}{}

\bibitem[Adelberger et al. 2004]{adelberger04} Adelberger, K. L.,
Steidel, C. C., Shapley, A. E., et al. 2004, ApJ, 607, 226

\bibitem[Bertin \& Arnouts 1996]{sex} Bertin, E. \& Arnouts, S.
1996, \aaps, 117, 393

\bibitem[Boulade et al. 2003]{boulade}
Boulade, O., Charlot, X., Abbon, P., et al. 2003, SPIE, 4841, 72

\bibitem[Capak et al. 2004]{capak04}
Capak, P., Cowie, L. L., Hu, E. M., et al. 2004, AJ, 127, 180

\bibitem[Eliche-Moral et al. 2006]{goya}
Eliche-Moral, M. C., Balcells, M., Prieto, M., et al. 2006, ApJ, 639, 644

\bibitem[Galadi-enriquez et al. 2000]{galadi00}
Galadi-enriquez, D., Trullols, E., Jordi, C. 2000, A\&AS, 146, 169 

\bibitem[Hill et al. 2000]{hill}
Hill, J. M., et al. 2000, in Science with the Large Binocular Telescope,
ed. T. Herbst (Heidelberg:Neumann Druck)

\bibitem[Landolt et al. 1992]{landolt92}
Landolt, A. U. 1992, AJ, 104, 372

\bibitem[Metcalfe et al. 2001]{wht}
Metcalfe, N., Shanks, T., Campos, A., McCracken, H. J. \& Fong, R. 2001,
MNRAS, 323, 795

\bibitem[Miyazaki et al. 2002]{subaru}
Miyazaki, S., et al. 2002, PASJ, 54, 833

\bibitem[Pedichini et al. 2003]{pedik}
Pedichini, F., Giallongo, E., Ragazzoni, R., et al.
2003, Proc. SPIE, 4841, 815

\bibitem[Pedichini \& Speziali 2004]{pedik04}
Pedichini, F., Speziali, R. 2004, ASSL, 300, 349

\bibitem[Radovich et al. 2004]{radovich}
Radovich, M., Arnaboldi, M., Ripepi, V. et al., 2004, A\& A, 417, 51

\bibitem[Ragazzoni et al. 2000]{ragazzoni}
Ragazzoni, R., et al. 2000, Proc. SPIE, 4008, 439

\bibitem[Ragazzoni et al. 2006]{ragazzoni06}
Ragazzoni, R., et al. 2006, Proc. SPIE, 6267, 33

\bibitem[Sawicki \& Thompson 2005]{sawicki05}
Sawicki \& Thompson 2005, ApJ, 635, 100

\bibitem[Speziali et al. 2004]{speziali}
Speziali, R., Pedichini, F., Di Paola, A., et al. 2004, SPIE, 5492, 900

\bibitem[Steidel et al. 1999]{steidel99} Steidel, C. C., Adelberger, K. L.,
Giavalisco, M., Dickinson, M., Pettini, M., 1999, ApJ, 519, 1

\bibitem[Steidel et al. 2003]{steidel}
Steidel, C. C., Adelberger, K. L., Shapley, A. E., et al. 2003, ApJ, 592, 728

\bibitem[Stone 1997]{stone}
Stone, R. C., 1997, AJ, 114, 2811

\bibitem[Veron-Cetty \& Veron 2006]{veron}
Veron-Cetty M.P., Veron P., 2006, \aap, 455, 773

\bibitem[Wynne 1996]{wynne}
Wynne, C. G., 1996, MNRAS, 280, 555

\end{thebibliography}
\end{document}